\documentclass[aps,prl,10pt,showpacs,amsmath,amssymb,floatfix,superscriptaddress,longbibliography,twocolumn]{revtex4-2}

\usepackage{physics}
\usepackage{physics}
\usepackage{graphicx}
\usepackage[dvipsnames,svgnames,table]{xcolor}
\usepackage{tcolorbox}
\usepackage{amsmath} 
\usepackage{amsfonts} 
\usepackage{amssymb} 
\usepackage{subfigure}
\usepackage{textcomp}
\usepackage{siunitx}
\usepackage[normalem]{ulem}
\usepackage[unicode=true,
pdfusetitle,
bookmarks=true,
bookmarksnumbered=false,
bookmarksopen=false,
breaklinks=true,
pdfborder={0 0 0},
backref=false,
colorlinks=true,
hypertexnames=false]{hyperref}
\hypersetup{linkcolor=NavyBlue,urlcolor=NavyBlue,citecolor=NavyBlue}
\usepackage[english]{babel}
\usepackage{bbm}
\usepackage{graphicx,color}

\definecolor{mygreen}{rgb}{0,0.5,0}
\definecolor{mygrey}{rgb}{0.5,0.5,0.5}
\definecolor{myred}{rgb}{0.75,0,0}
\definecolor{myblue}{rgb}{0,0,0.75}
\definecolor{mymagenta}{cmyk}{0,1,0,0.12}
\definecolor{mycyan}{cmyk}{1,0,0,0.12}
\definecolor{myorange}{rgb}{1,0.5,0}
\definecolor{myviolet}{rgb}{0.5,0.0,0.75}
\definecolor{mybrown}{cmyk}{0,0.50,1,0.41}
\definecolor{mygrey}{rgb}{0.5,0.5,0.5}

\newcommand{\Rb}{\textsuperscript{87}Rb~}
\newcommand{\myaffiliation}{\affiliation}
\newcommand{\HDU}
{\myaffiliation{Department of Physics, Hangzhou Dianzi University, Hangzhou 310018, China}}
\newcommand{\ZKL}
{\myaffiliation{Zhejiang Key Laboratory of Quantum State Control and Optical Field Manipulation, Hangzhou Dianzi University, Hangzhou 310018, China}}
\newcommand{\ICFO}
{\myaffiliation{ICFO - Institut de Ci\`encies Fot\`oniques, The Barcelona Institute of Science and Technology, 08860 Castelldefels (Barcelona), Spain}}
\newcommand{\ICREA}{\myaffiliation{ICREA - Instituci\'{o} Catalana de Recerca i Estudis Avan{\c{c}}ats, 08010 Barcelona, Spain}}

\begin{document}    
	\title{Adaptive Real-Time Magnetic Field Tracking beyond Prior Waveform Constraints \\
}
    \author{Yihan Wang}
    \HDU
    
	\author{Xiaofeng Jin}
    \HDU
    
	\author{Yuchuan Ming}
    \HDU
    
    \author{Jianxiang Miao}
    \HDU
    
	\author{Xiao-Ming Lu}
    \email{lxm@hdu.edu.cn}
    \HDU
    \ZKL

    \author{M. W. Mitchell}
    \ICFO
    \ICREA
    
    \author{Jia Kong}
	\email{jia.kong@hdu.edu.cn}
    \HDU
    \ZKL
	\begin{abstract}
		The extraction of weak signals plays a crucial role in quantum precision measurement, where the estimation results are often limited by low signal-to-noise ratios. Here, we demonstrate a parameter-estimation framework based on the adaptive extended Kalman filter for dynamic magnetic-field estimation in quantum systems using spin-noise measurements---a challenging regime characterized by weak signals. By modeling the magnetic field as an unknown parameter, the proposed approach alleviates model dependence in state estimation. Furthermore, by introducing an adaptive algorithm with real-time noise estimation, our method overcomes the measurement noise intensity constraints of conventional extended Kalman filtering and enhances its practical applicability. Numerical simulations covering three representative magnetic-field dynamics validate the capability of the proposed framework, while experimental results demonstrate successful tracking of a seismo-magnetic-like signal beyond the intrinsic sensitivity of conventional spin-noise spectroscopy.
	\end{abstract}
	\maketitle
    
    The extraction of weak signals is a central task in precision measurement, with important applications including phase estimation, gravitational wave detection~\cite{PhysRevD.81.084029,RevModPhys.94.025001}, and fundamental physics research~\cite{PhysRevLett.97.131801,PhysRevLett.111.102001}. Although the information is encoded in the measurement record, reliable reconstruction becomes difficult as the signal approaches the noise floor. This challenge is particularly severe in noise-based measurements, where the signal of interest must be extracted from intrinsic fluctuations, such as in the search for space-time fluctuations associated with quantum gravity~\cite{PetruzzielloNatureCommunications2021}, in which the inherently low signal-to-noise ratio (SNR) imposes fundamental limits on parameter estimation accuracy. Magnetic field estimation from spin noise is a representative example: the extremely low SNR renders commonly-employed signal extraction highly nontrivial, and signal accumulation via spin noise spectroscopy (SNS) based on Faraday rotation is commonly used~\cite{Kong2020,Crooker2004,Zapasskii2013,Zapasskii2013a}. SNS accesses intrinsic spin fluctuations through frequency-domain analysis of voltage time series~\cite{CohenTannoudji1969} and has been successfully applied to atomic spin dynamics~\cite{Crooker2004,Katsoprinakis2007}, conduction electron systems~\cite{Crooker2010}, and localized states in semiconductors~\cite{Oestreich2005}. However, as a steady-state frequency-domain method, SNS provides only time-averaged magnetic-field information and is limited in resolving fast or time-dependent field dynamics.
    
    
    The Kalman filter (KF)~\cite{Kalman1960,Kalman1961} is particularly proficient at processing time-series signals and tracking dynamic parameters/state variables, making it well-suited to overcome the above limitations. As a Bayesian estimator, it offers low computational cost, broad applicability, and the ability to infer unobservable physical variables, with established use in inertial navigation~\cite{Grewal2020}.  More recently, KF has been successfully applied in quantum systems and precision metrology~\cite{JimenezMartinez2018,Ma2022,AmorosBinefa2021,k7nk-lrwd,amorosbinefa2025trackingtimevaryingsignalsquantumenhanced,Emzir2017,Geremia2003} for real-time tracking and parameter estimation. However, its performance strongly depends on accurate system modeling, particularly the state-space dynamics and noise statistics~\cite{Revach2022,article,Mohamed1999a}.

    \begin{figure*}[!]
		\includegraphics[width=0.98\textwidth]{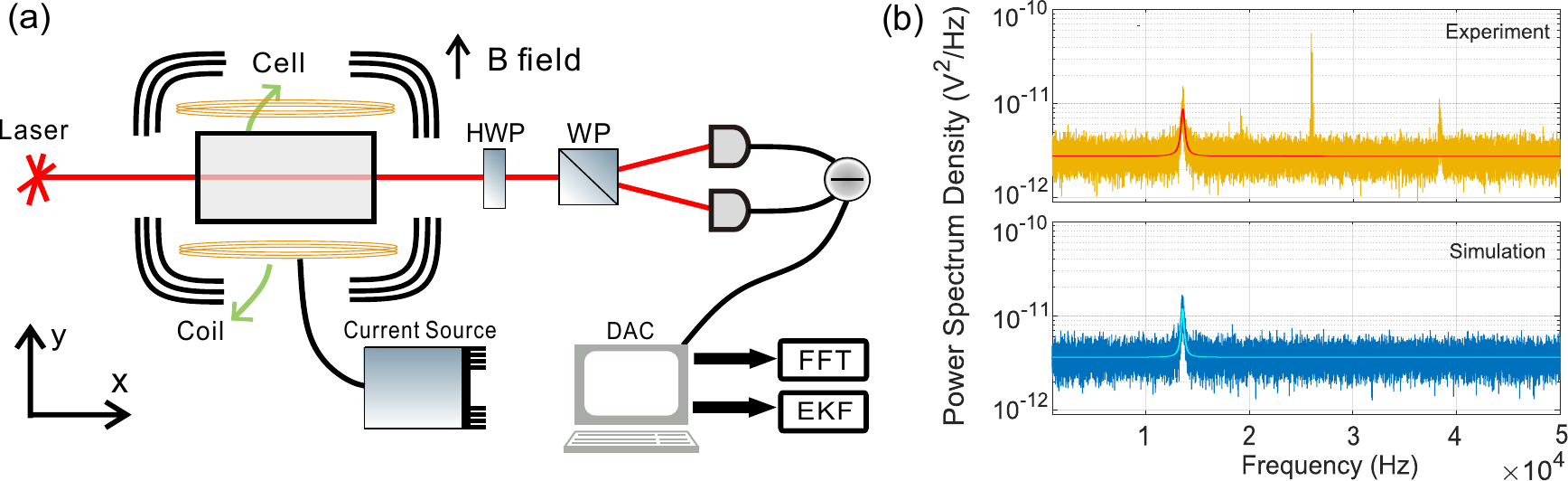}
		\centering
		\caption{\label{aa}(a)Experimental schematic. The probe light propagates along x-direction, perpendicular to magnetic field oriented along y-direction, through a \Rb atomic cell. The atoms precess at the Larmor frequency ${\omega}$, which is set by current-controlled external magnetic field. According to the Faraday effect, spin fluctuations induce $S_2$ Stokes component fluctuations in the probe light, detected by a balanced photodetector and processed by a data acquisition card (DAC). HWP: half wave plate, WP: Wollaston prism. (b)Spin noise spectra (SNS) obtained from both experiment and simulation via fast Fourier transform (FFT), with light blue and red lines representing the respective fitting curves. }
	\end{figure*}    
    
    To address these challenges, we treat the magnetic field as an unknown parameter rather than a state variable~\cite{Verstraete2001}, enabling estimation of arbitrarily time-varying fields without assuming a specific dynamical model. In addition, to handle non-stationary and difficult-to-characterize experimental noise, we employ an adaptive Kalman filter (AKF)~\cite{Mehra1970} that updates the measurement noise covariance in real time, reducing manual tuning and improving robustness in quantum metrology. We demonstrate the effectiveness of this approach in atomic spin-noise–based magnetic-field estimation, a challenging regime where suboptimal conditions limit precision. Numerical simulations show accurate tracking of dynamic fields without prior knowledge of either the signal model or noise strength, while experiments further confirm its practical feasibility. Overall, this work links estimation theory with quantum metrology and highlights the potential of Kalman filtering for precision physics applications.

    
    The experimental setup, shown in Fig.~\ref{aa}(a), features a \Rb atomic vapor cell at the center of magnetic shields, with the target field, $\textit{B(t)}$, applied along the $y$ direction. The atoms are initialized in an unpumped thermal equilibrium state, creating a measurement scenario characterized by strong noise and weak signals. The subsequent dynamical evolution of the atomic spin components is governed by the Bloch equations~\cite{Seltzer2008,Bloch1946}:    
    \begin{equation}\label{1}
		d\mathbf{J}_t=
		\begin{bmatrix}
			-\Gamma & -\omega(t)\\
			\omega(t) & -\Gamma
        \end{bmatrix}
        \mathbf{J}_t \,dt+\sqrt{2\Gamma\mathbf{q}}\,d\mathbf{W}_t,
    \end{equation}
    where the spin vector $\mathbf{J}_t = [J_{x,t}, J_{z,t}]^T$ follows linear dynamics~\cite{PhysRevA.93.053802,PhysRevA.95.041803}, $\Gamma$ denotes relaxation term, $\mathbf{q}=\operatorname{diag}\left \{ q_{x}, q_{z} \right \} $ is the spin-noise fluctuation matrix and $d\mathbf{W}_t = [dW_x, dW_z]^T$ denotes the spin noise increment, which obeys Gaussian white-noise statistics. Here, ${\omega}(t) = {\gamma}B(t)$ denotes the Larmor precession frequency, with $\gamma$ being the gyromagnetic ratio. 

    Consequently, the magnetic field information embedded in the spin noise is detected through Faraday rotation. The resulting polarization fluctuations are then sensed by a balanced photodetector (BPD) and converted into photocurrent:
	\begin{equation}\label{2}
		I_{t} = g_{D}J_{x,t}+V_{t},
	\end{equation}
    where $g_D$ is the coupling constant between probe light and atomic spin, and $V_{t}{\sim}\mathcal{N}(0,R_{t})$ represents the Gaussian-distributed measurement noise introduced during the detection process. This noise satisfies $\textit{E}[V_{t}V_{j}x^{T}]=R_{t}{\delta}_{tj}$ where $R_{t}$ denotes its intensity, which can be estimated from the power spectral density~(PSD). 
    
    The model, built upon Eq.~\eqref{1} and Eq.~\eqref{2}, enables numerical simulation via the Monte-Carlo method. The simulation parameters are chosen according to the experimental conditions: a cylindrical \Rb vapor cell of length $L_\text{cell}=3$ cm, atomic density $n=1.37{\times}10^{13}~\text{cm}^{-3}$, 100 Torr of $\mathrm{N}_2$ buffer gas, and a temperature of $110~^\circ\text{C}$ controlled by a thermostat. The probe laser is red-detuned by 16 GHz from the $D_1$ line at $\nu_0=377{.}111{\times}10^{12}$ Hz, with a power of $P=1$ mW over an effective area of $2.66~\text{mm}^2$. The detector parameters are: quantum efficiency $\eta=0.87$, transimpedance gain $G=10^{6}$ V/A, and sampling rate $F_s=200$ kSa/s. The photocurrent signals from both experiment and simulation were processed using Fast Fourier Transform (FFT) to yield SNS signals in Fig.~\ref{aa}(b). These spectra were fitted with a Lorentzian profile \cite{Sinitsyn2016}. The agreement between the experimental and simulation results confirms that our simulation can be reliably used to evaluate tracking performance.

	\begin{figure*}[t]
 		\includegraphics[width=0.98\textwidth]{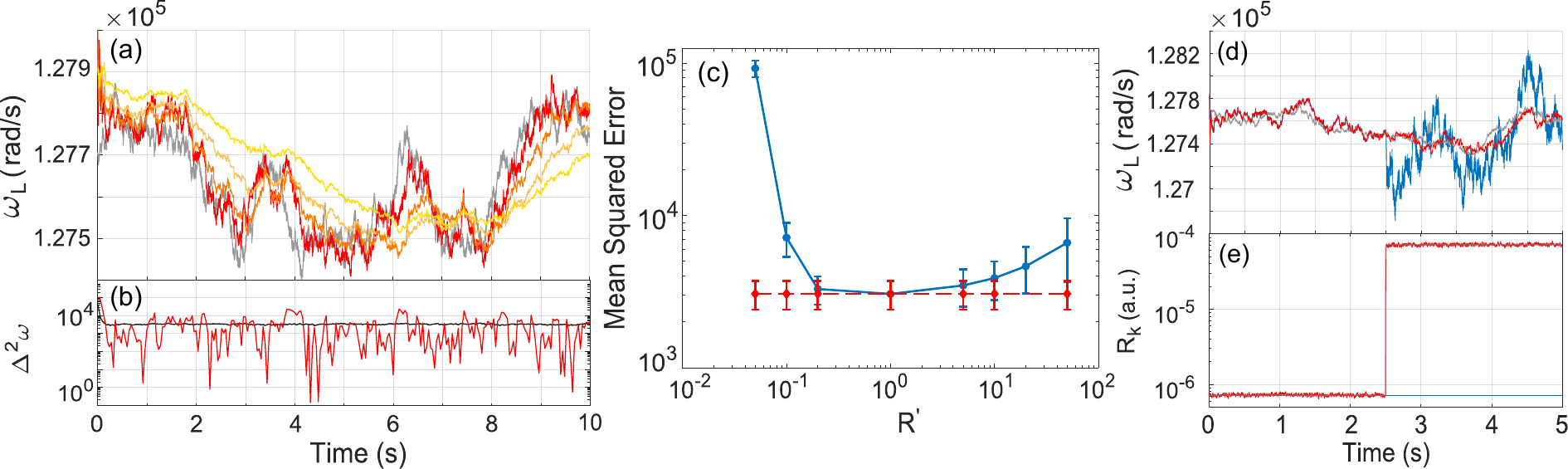}
		\centering
		\caption{\label{bb} (a) Larmor frequency $\omega$ generated by Eq.\eqref{eq:OmegaUpdate} with $\sigma^2 = 0.1$ as the ground truth (gray solid line), together with the corresponding AEKF estimation result (red solid line) with measurement-noise deviation $R' = 1000$. Red-orange, orange and yellow traces show EKF estimates with progressively increased measurement-noise deviations, with values of $R' = 50$, $100$, and $200$, respectively.
        (b) Squared instantaneous estimation error ($\Delta^{2}\omega$, red solid line) and the variance estimated by the AEKF (black solid line).
        (c) Mean squared error (MSE) as a function of the estimation performance used in the EKF and AEKF. We evaluate the average instantaneous variance of simulation data and calculate the mean and error based on 50 sets of data. The blue and red curves correspond to the EKF and AEKF results, respectively.
        Larmor frequency estimation (d) and adaptive measurement-noise identification (e) under environmental interference. The gray curve shows the true value, while the red and blue curves represent the EKF and AEKF estimates. At $t=2.5~\mathrm{s}$, the measurement noise increases abruptly.}
	\end{figure*}
   
    As a frequency-domain analysis, SNS inherently provides only time-averaged magnetic field measurements. Achieving sufficient SNR requires extensive FFT-based signal accumulation, limiting temporal resolution. However, from a time-domain perspective, commonly-employed signal extraction is hindered by extremely weak signals from unpolarized samples. We therefore employ a KF for real-time variables estimation in a magnetically driven spin system, enabling efficient dynamic tracking. We note that, for magnetic field estimation beyond spin-state estimation, it is necessary to augment the state matrix~\cite{Verstraete2001} and model the parameter $\omega$ as a slowly time-varying variable. This leads to nonlinear system dynamics and motivates formulation of the extended Kalman filter (EKF) ~\cite{BELLANTONI1967}, with first-order Taylor series expansion remaining accurate for weak nonlinearities.
    
    Solving and discretizing Eqs.\eqref{1} and \eqref{2} yields the following state and measurement vector model:
	\begin{align}
		\mathbf{x}_{k}&=\mathbf{f}_{k-1}(\mathbf{x}_{k-1})+\mathbf{w}_{k-1},\label{3}\\
		y_{k}&=\mathbf{H}_{k}\mathbf{x}_{k}+V_{k}\label{5},
	\end{align}	
	where $\mathbf{H}_{k}=[g_{D},0,0]$, $\mathbf{x}_{k}=[J_{x,k},J_{z,k},\omega_{k}]^{T}$,
 \begin{multline*}
        \mathbf{f}_{k-1}(\mathbf{x}_{k-1})= \\
        \begin{bmatrix}
			e^{-\Delta\Gamma}(\cos({\omega}_{k-1}{\Delta})J_{x,k-1}-\sin({\omega}_{k-1}{\Delta})J_{z,k-1})\\
			e^{-\Delta\Gamma}(\sin({\omega}_{k-1}{\Delta})J_{x,k-1}+\cos({\omega}_{k-1}{\Delta})J_{z,k-1})\\
			{\omega}_{k-1}
		\end{bmatrix},
\end{multline*}
and ${\Delta}=t_{k+1}-t_{k}$ denotes the discrete-time interval, set equal to the sampling interval of detector and $k$ is an integer represents discrete sampling time. The process noise vector is given by $\mathbf{w}_{k-1}=[w_{x,k-1},w_{z,k-1},w_{p}]^{T}$, which comprises Wiener increments. The process noise intensities are modeled as $w_{s, k} \sim \mathcal{N}(0,\left ( 1-e^{-2\Gamma \Delta }  \right ) q_{s})$ for $s \in {x,z}$, where $q_{s}$ denotes the spin-noise fluctuation. Their cross-covariance is specified by $\textit{E}[w_{x,k}w_{z,j}^{T}]=\left ( 1-e^{-2\Gamma \Delta }  \right ) q_{s}{\delta}_{xz}{\delta}_{kj}$~\cite{Kong2020,JimenezMartinez2018}. The term $w_{p} \sim \mathcal{N}(0,\alpha)$ is an artificial noise with a small intensity $\alpha$, introduced to drive the estimation of the unknown parameter~\cite{Simon2006}. According to the process equation, the state estimate and estimation-error covariance in the time-update step are calculated using the EKF.
 	 
    
    However, the performance of EKF hinges on accurate dynamic and noise models. While the system dynamics can be well-defined in practice, precise noise modeling remains a significant challenge, which can severely degrade the filter's performance. To overcome both the challenges in  noise estimation and its potential non-stationarity, we incorporate the Sage-Husa adaptive algorithm into the EKF framework, resulting in the adaptive extended Kalman filter (AEKF) ~\cite{Mohamed1999}.

	The fundamental principle of the AEKF lies in dynamically updating measurement noise through innovation-based adaptive estimation~(IAE) with a sliding window,
	\begin{align}
		\hat{R}_{k} &=\mathbf{H}_{k}\hat{\mathbf{P}}_{k}\mathbf{H}_{k}^{T}-\frac{1}{N}\sum_{j=0}^{N}v_{k-j}v_{k-j}^{T},\label{14}\\
		v_{k} &=z_{k}-\mathbf{H}_{k}\hat{\mathbf{X}}^{-}_{k},\label{15}
	\end{align}
	where $\hat{R}_{k}$ denotes adaptively estimated measurement noise, $N$ denotes the length of estimation window and $v_{k}$ denotes innovation sequence at time $k$. 
    In contrast to the EKF, Eqs.~\eqref{14} and \eqref{15} introduce only a marginal computational overhead to the algorithm, while eliminating the need for empirical dependence on measurement noise.
    
    The simulation allows arbitrary definition of magnetic field variations while retaining complete knowledge of the true field values at every time $t_k$, enabling straightforward estimation evaluation. In the simulation process, we model the unknown and randomly fluctuating magnetic field using a random walk process---a Markovian model suitable for unordered stochastic variations --- as described by:
	\begin{equation}
    \label{eq:OmegaUpdate}
		{\omega}_{k+1}={\omega}_{k}+d{\omega},
	\end{equation}
    where $\omega_{k}$ is Larmor frequency generated by simulation program at time $t_k$, $d\omega \sim \mathcal{N}(0, \sigma^2)$ denotes the stochastic independent increment and $\sigma$ represents the standard deviation of the increment.
    
    Figure~\ref{bb} (a) shows one simulation process and corresponding estimation results. The measurement noise mismatch is defined by the ratio $R' \equiv R_1 / R_{\text{true}}$, where $R_{\text{true}}$ denotes the true measurement noise intensity determined by shot noise~\cite{JimenezMartinez2018} and remains constant throughout the simulation, and $R_1$ is the assumed measurement noise intensity in EKF/AEKF at time $t_1$. In the EKF, this parameter keep constant during the entire estimation process, whereas the AEKF adaptively updates the measurement noise intensity in real time according to Eqs.~\eqref{14}. A progressive increase in measurement noise deviation is visually represented by a color gradient transitioning from orange to yellow, which correlates with a corresponding degradation in estimation accuracy of the EKF. In contrast, the AEKF estimate (red), initialized with a measurement noise strength that deviates significantly from the true value ($R'=1000$), closely tracks the ground truth (gray). The performance of AEKF remains robust across different degrees of measurement-noise mismatch and is comparable to that of the optimal EKF estimation. 
    Moreover, the agreement between the squared estimation error ${\Delta}^{2}{\omega}_{k}=({\omega}_{k}-\hat{{\omega}_{k}})^{2}$ and the predicted covariance $\hat{P}_{k}$, shown in Fig.~\ref{bb}(b), verifies that the AEKF provides a reliable error estimate throughout convergence.

    \begin{figure}[!]
		\includegraphics[width=0.49\textwidth]{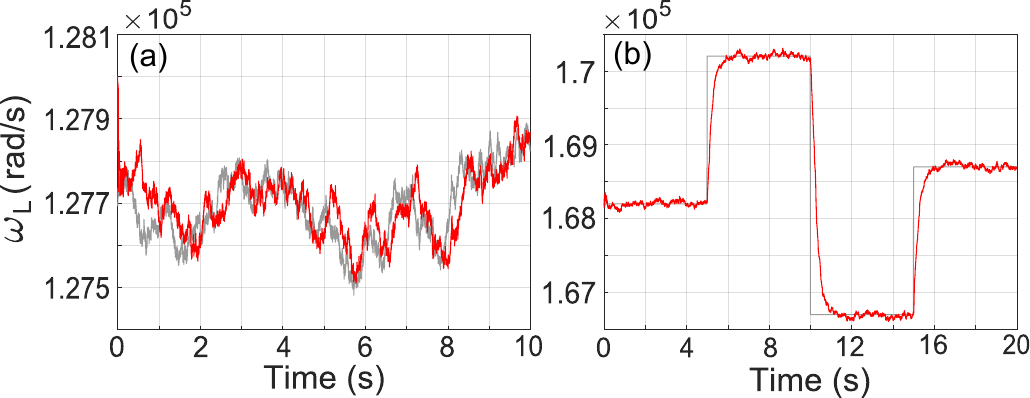}
		\centering
		\caption{\label{dd}AEKF-based estimation results for magnetic fields with unknown temporal variations. The true magnetic-field dynamics (gray solid line) and the corresponding AEKF estimates (red solid line) are shown for two representative models: (a) an Ornstein–Uhlenbeck process and (b) a piece-wise constant field, demonstrating good agreement between the estimates and the true dynamics.}
	\end{figure}
    
    To quantify the impact of noise model mismatch, we evaluate the estimation error under different deviations ($R'$). The mean squared error (MSE), shown in blue in Fig.~\ref{bb}(c), is computed from 50 independent EKF runs, excluding the initial 0.1 s to ensure steady-state analysis. Results show that both underestimated ($R' < 1$) and overestimated ($R' > 1$) noise covariance lead to degraded tracking due to excessive fluctuations or insufficient response, causing EKF failure. This highlights a key limitation in practical systems where time-varying noise is difficult to model and may fluctuate due to environmental coupling. In contrast, the MSE of AEKF (red in Fig.~\ref{bb}(c)) remains stable at $3059~(rad/s)^2$ across all $R'$, close to the EKF benchmark of $3055~(rad/s)^2$ at $R'=1$. The results demonstrate that the AEKF achieves accurate estimation even without an accurate noise model, which gives robustness in fidelity in practical applications.


    Figures~\ref{bb}(d) and (e) demonstrate the robustness of the AEKF under environmental disturbances. The magnetic field follows a random-walk model over 5~$\mathrm{s}$, with initial EKF and AEKF noise settings matching the true value. At $t=2.5~\mathrm{s}$, the measurement noise intensity increases abruptly by two orders of magnitude due to external disturbance. As shown in Fig.~\ref{bb}(d), both filters perform comparably before the disturbance which accurately tracking the magnetic-field evolution. Afterwards, the EKF rapidly loses tracking capability due to measurement-noise mismatch, whereas the AEKF continues to stably track the magnetic field. Figure~\ref{bb}(e) shows that the AEKF adaptively updates the observation-noise covariance via the innovation sequence, enabling rapid identification of noise variations and preventing divergence.
    
    In addition to the random-walk field, we also simulate the Ornstein–Uhlenbeck (OU) process and piece-wise constant variation, as shown in Fig.~\ref{dd}(a) and (b), respectively, to evaluate the performance of the proposed method for magnetic-field estimation in the absence of prior knowledge. It is worth noting that although the state equation in Eqs.~\eqref{3} remains unchanged across all tested cases, the AEKF successfully tracks the different field evolutions in real time, with the estimates remaining in close agreement with the true values throughout most of the evolution. These results highlights the key advantage that the proposed method can robustly reconstruct time-domain magnetic field variations without prior knowledge of the dynamics equation.

	\begin{figure}[t]
		\includegraphics[width=0.48\textwidth]{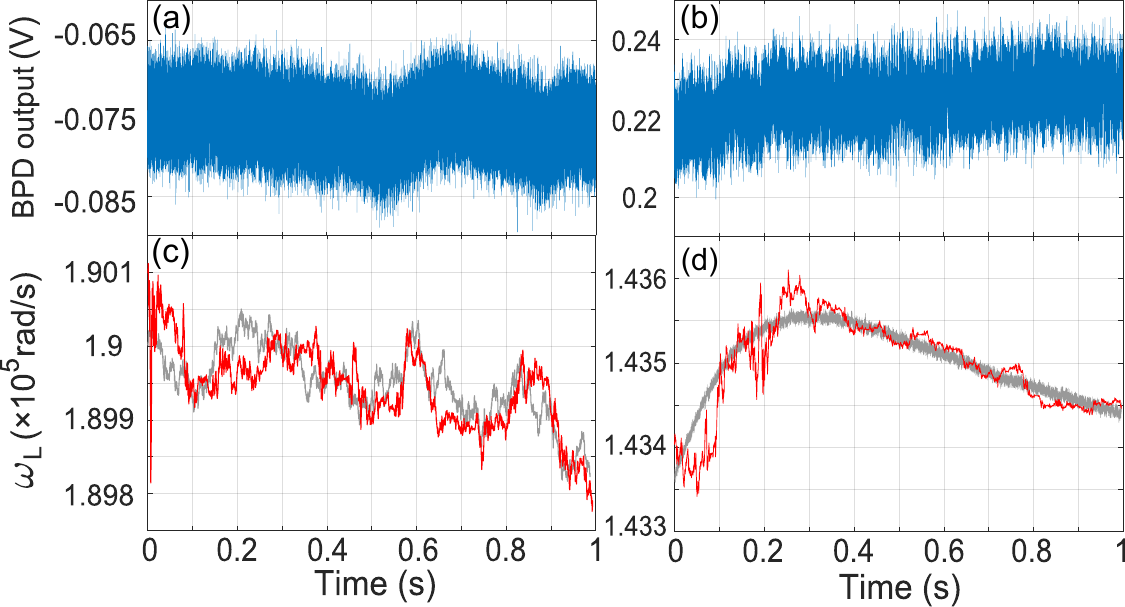}
		\centering
		\caption{\label{ee} Experiment results.
        (a)-(b) Faraday rotation observations detected by BPD with a random-walk and SE-like magnetic field, respectively. Under such low-SNR conditions, the time-domain magnetic signal is obscured by noise, making 
        commonly-employed time-domain fitting challenging.
        (c)-(d) Random-walk and SM-like magnetic field generated by the signal generator (gray solid line) respectively, together with the corresponding AEKF estimates (red solid line) obtained without prior knowledge of the field-evolution model.}
	\end{figure}
    
    Building on the above results, we further validate the proposed tracking method using experimental data from an $^{87}$Rb vapor cell, as shown in Fig.~\ref{1}(a). In experiments, a random-walk magnetic field and a seismo-magnetic~(SM) like field are used to emulate realistic field drifts and earthquake-induced magnetic perturbations, respectively. As shown in Fig.~\ref{ee}(a)-(b), the measured voltage signal is strongly contaminated by noise, making conventional fitting methods~\cite{Kong2025} and SNS-based extraction ineffective. In contrast, as shown in Fig.~\ref{ee}(c), the AEKF successfully reconstructs random-walk magnetic field variations on the order of 0.22 nT from spin-noise measurements. The reconstructed field remains below the SNS sensitivity limit ($18.57nT/\sqrt{Hz}$), experimentally determined from the fitted spin-noise spectrum under a static magnetic field using the sensitivity definition in Ref.~\cite{Groeger2006}. This indicates effective noise suppression by the AEKF and an associated improvement in sensitivity. For SM monitoring, we generate a magnetic field signal based on the 1998 San Andreas Fault earthquake event, where magnetic anomalies may accompany seismic activity due to crustal fracturing processes~\cite{rs14225893,VAROTSOS1986,HAYAKAWA2004617,VAROTSOS19931}. Such signals are inherently nonstationary and unpredictable, making model-free estimation essential. The AEKF estimates are shown in Fig.~\ref{ee}(d), which accurately tracks the underlying dynamics, successfully recovering the weak magnetic-field signal. A short convergence period is required due to the absence of prior knowledge, performance would further improve if a precise field model were available. However, for practical applications such as real-time seismic magnetic monitoring, such prior models are generally unavailable. Consequently, the proposed parameter-estimation approach offers a more realistic and broadly applicable solution.
    
	In conclusion, we have demonstrated the feasibility of employing AEKF for dynamic magnetic field estimation in quantum systems, establishing its ability to extract weak magnetic field signals under low SNR. By formulating the magnetic field as an unknown parameter, we achieve the estimation of arbitrarily varying magnetic fields without prior knowledge of their dynamical evolution model. The incorporation of innovation-based adaptive algorithm further enables real-time measurement noise covariance adaptation, effectively eliminating empirical parameter dependence. This work establishes AEKF as a promising framework for noise-dominated parameter estimation and weak-signal reconstruction in precession measurements.

	We acknowledge support from the Quantum Science and Technology-National Science and Technology Major Project (Grant No. 2024ZD0302200, No. 2024ZD0301000), and National Natural Science Foundation of China (NSFC) (Grants No.62522504, No.12374463, No.92476118 and No.12275062). This work is also supported by the key R$\&$D Program of Zhejiang (2026C01004), Zhejiang Provincial Natural Science Foundation of China (ZCLQN25A0404) and Fundamental Research Funds for the Provincial Universities of Zhejiang (No. GK249909299001-002). MWM supported by  European Commission projects Field-SEER (ERC 101097313) and QUANTIFY (101135931); Spanish Ministry of Science MCIN projects SAPONARIA (PID2021-123813NB-I00) and SALVIA (PID2024-158479NB-I00),  ``Severo Ochoa'' Center of Excellence CEX2024-001490-S [MICIU/AEI/10.13039/501100011033];  Generalitat de Catalunya through the CERCA program,  DURSI grant No. 2021 SGR 01453 and QSENSE (GOV/51/2022).  Fundaci\'{o} Privada Cellex; Fundaci\'{o} Mir-Puig.

	\bibliography{wyhrefe}
\end{document}